\documentclass[structabstract]{aa}  
\usepackage{graphicx}
\usepackage{txfonts}
\begin{document}
   \title{VLT Phase Resolved Optical Spectroscopy of the
Ultra-Compact Binary HM~Cnc}

   \subtitle{}

   \author{Elena Mason\inst{1}
	  G.L. Israel\inst{2}
	  S. Dall'Osso\inst{2,3}
	  L. Stella\inst{2}
	  U. Munari\inst{4}
	  G. Marconi\inst{1}
	  K. O'Brian\inst{1}
	  S. Covino\inst{5}
          \and
	  D. Fugazza\inst{5}
          }

   \institute{ESO - European Southern Observatory, Alonso de Cordova 3107, Vitacura, Santiago, Chile\\
   \email{emason@eso.org}
   \and
INAF - Osservatorio Astronomico di Roma, Via Frascati 33, I-00040 Monte Porzio Catone, Italy\\
    \email{gianluca@oa-roma.inaf.it}
\and 
	  Dipartimento di Fisica 'Enrico Fermi', Universit\a di Pisa, Largo B. Pontecorvo 3, I-56127
Pisa, Italy
\email{dallosso@oa-roma.inaf.it}
\and
INAF - Osservatorio Astronomico di Padova, Sede di Asiago, I-36012 Asiago, Italy
\and 
INAF - Osservatorio Astronomico di Brera, Via Bianchi 46, I-23807 Merate, Italy
             }

   \date{Submitted in April 2008; Updated text after the second referee report}

 
  \abstract
  {A 321.5 s modulation was discovered in 1999 in the X-ray light curve of HM~Cnc.
In 2001 and 2002, optical photometric and spectroscopic observations revealed that HM~Cnc is a very
blue object with no intrinsic absorptions but broad (FWHM$\sim$1500 km s$^{-1}$) low equivalent width
 emission lines (EW$\sim1\div6$\AA), which were first identified with the HeII Pickering series. The
combination of X-ray and optical observations pictures HM~Cnc as a double degenerate binary
 hosting two white dwarfs, and possibly being the shortest orbital period binary discovered
so far.} 
   {The present work is aimed at studying the orbital motion of the two components by following
the variations of the shape, centroid and intensity of the emission lines through the orbit.}
   {In February 2007, we carried out the first phase resolved optical spectroscopic study
with the VLT/FORS2 in the High Time Resolution (HIT) mode, yielding five phase bins in the
321 s modulation. }
   {Despite the low SNR, the data show that the intensity of the three most prominent emission
lines, already detected in 2001, varies with the phase. These lines are detected at phases 0.2-
0.6 where the optical emission peaks, and marginally detected or not detected at all elsewhere.
Moreover, the FWHM of the emission lines in the phase resolved spectra is smaller, by almost a factor 2, than that in the the phase-averaged 2001 spectrum.}
   {Our results are consistent with both the pulsed optical component and emission
lines originating in the same region which we identify with the irradiated surface of the secondary.
Moreover, 
we note that the
EWs of the emission lines might be up to $-15 \div -25$ \AA, larger than thought before; these values
are more similar to those detected in cataclysmic variables. All the findings further confirm that the 321s modulation observed in HM~Cnc is the orbital period of the system,
the shortest known to date.}
   {}

\authorrunning{Mason E. et al.}
\titlerunning{First HM~Cnc time resolved spectroscopy}
   \maketitle
%


\section{Introduction}

Double degenerate binaries (DDBs) are close binary star systems consisting of two white dwarfs
(WD) or a WD and a He star. DDBs are expected to originate from main sequence stars which have experienced two common
envelope phases, eventually exposing the CO or He cores of the original stars (for a review
see Warner 1995). Among DDBs are AM CVns, systems in which either a low mass He~WD or a low mass non degenerate He burning star fills its Roche lobe and transfers matter to the companion (Nelemans et al. 2001a). AM CVn binaries are characterized by orbital
periods in the 600-3000 s range and, in the case of non degenerate donors, they evolve from long to short orbital periods (till they reach the minimum orbital period of $\sim$10 min), similarly to cataclysmic variables (CVs). They evolve toward longer orbital periods, otherwise (Nelemans et al. 2001a and reference therein). 
Gravitational radiation waves drive the mass transfer and double-peaked emission lines in the optical spectra of most AM CVns testifies to the presence of an accretion disk mediating the flow of matter from one star to the other.

In 1999, Israel et al. (1999, hereafter I99) discovered an X-ray emitting source with a strong
321.5~s. Deep TNG and VLT observations
were carried out to study the optical counterpart of HM~Cnc and its optical spectral features.
HM~Cnc is a blue V=21.1 (B=20.7) star showing a $\sim$15\% pulsed fraction at the $\sim$321.5 s X-ray period
(Israel et al. 2002, hereafter I02; see also Ramsay et al. 2002a). The optical and X-ray pulsations
displays a phase offset of $\sim$0.20-0.25 (Barros et al. 2007) indicating that they originate from two different regions. The
optical modulation also varied in shape with time, while its pulsed fraction was clearly wavelength dependent
(being larger at longer wavelengths). The spectrum of HM~Cnc is characterized by a blue
continuum with no intrinsic absorption lines (I02), but broad (FWHM$\sim$1500 km s$^{-1}$), low equivalent
width (EW $\sim$1-6 \AA) emission lines, which were first identified as lines from the HeII Pickering
series.
All these findings are fully compatible with HM~Cnc being an X-ray emitting, He-rich DDB,
possibly a progenitor of the AM CVn systems, with the shortest orbital period ever recorded (I02).
Due to the difference in EWs measured between the odd- and even-terms of the HeII Pickering
series, it has been proposed that HM~Cnc may also be hydrogen-rich (Steiper et al. 2005 and references
therein). Should this be confirmed, it would be diffcult to relate HM~Cnc with AM CVn systems
which show only He emission/absorption lines in their spectra.
Thanks to the 2001-2006 photometric monitoring, Israel et al. (2004) found a very accurate P
and $\dot{P}$ phase coherent solution, with $\dot{P}= -$1.1$\times$10$^{-3}$ s/yr implying that the orbit is decaying. The
optical solution was sufficiently accurate that it could be extended backward in time so as to find a
unique phase-connected solution encompassing ROSAT 1994 and 1995 observations. This yielded
a 10-years baseline P-$\dot{P}$ coherent solution giving P=321.53038(2)~s and $\dot{P}=-$3.661(5)$\times$10$^{-11}$~s s$^{-1}$ 
(for more details see Israel et al. 2004; see also Strohmayer 2005 and Barros et al. 2007).
Yet, the nature of the X-ray emission detected from HM~Cnc and its twin source V407~Vul is
still under debate. A number of models have been proposed (for a review see Cropper et al. 2004).
Among these, is the DDB model with mass transfer proposed in two flavors: with a magnetic
primary (polar-like, Cropper et al. 1998) and with a non-magnetic accretor (Algol-like, Marsh \&
Steeghs 2002, Ramsay et al. 2002b). In the latter model, the reason why a disk does not form is because the minimum
distance of the gas stream from the center of mass of the system is smaller than the size of the
accretor, resulting in the stream hitting the surface of the accretor. The unipolar inductor model,
UIM, is a second accredited model and, according to it, the secondary star does not fill its Roche-lobe but, by moving 
across the primary star magnetic field, induces an electric current which heats up the B-field
line footprints on the primary white dwarf. No mass transfer is expected within the UIM. A similar mechanism
is likely responsible for the discovered Jupiter-Io interaction (Wu et al. 2002; Clarke et al. 1996;
Dall'Osso et al. 2006, 2007). In the DDB scenario the X-ray emitting region is thought to be ``self-eclipsed'' by the primary object giving rise to the sharp and large amplitude X-ray modulation. In
all models the source is expected to be one of the strongest sources of low frequency gravitational
wave radiation, easily detectable in the future by the LISA mission (Nelemans et al. 2001b and
references therein).
Within the DDB scenario, the negative value of the period derivative suggests that matter transfer
is not viable, therefore disfavoring the Polar-like and Algol-like scenarios, and the fact that the
GW emission is driving the orbital period decay. Moreover, the $\dot{P}$ value is slightly large for the
gravitational wave emission mechanism as the only channel governing the period evolution and
we cannot currently exclude that magnetic stresses are also present in the system (Dall'Osso et al.
2006, 2007). However, Deloye and Taam (2006) have revised AM~CVn formation and evolution, showing that fully degenerate donors and the accretion scenario can easily produce systems having the orbital periods of HM~Cnc and V407~Vul at early contact. In addition, they also revise the duration of the phase characterized by mass transfer and negative $\dot{P}$, determining values of $\tau_{\dot{M}}$ which are comparable with those predicted by the UIM. An alternative way around the negative $\dot{P}$ during accretion has been proposed by D'Antona et al.
(2006), who suggest that HM~Cnc is an accreting DDB, which is characterized by p-p hydrogen
shell burning in the donor, seen during a mass transfer phase. The nuclear reactions on the shell
cause the secondary star to shrink in response to the mass loss. D'Antona et al. (2006) obtain a
period derivative which is correct in both sign and magnitude. This scenario also accounts for the
possible presence of H in the spectrum of HM~Cnc. The lack of (relatively large) optical polarization
argues against the Polar-like scenario. We will not consider this model further in this paper
(see Reinsch et al. 2004; Israel et al. 2004). Recently, Ramsay et al. (2007) reported the detection
of transient radio emission from HM~Cnc, which can be accounted for only with maser emission in
the framework of the UIM model.

In 2006 we requested and obtained VLT-FORS2 observations in a new observing mode, aimed
at carrying out the first ever phase-resolved spectroscopic study of the 5.4 min modulation of HM~Cnc.
Our main goals included the disentangling of the different optical components (pulsed vs un-pulsed), the identification of the emission line forming region(s) and the detection of the emission lines Doppler modulation to provide the ultimate proof of HM~Cnc binary nature with orbital period OP=5.4~min. Here, we report on the results we obtained from this new spectroscopic campaign. In section 2 we describe the observation and the FORS2 HIT-S mode. In section 3 the obtained results, while in section~4 we report our conclusions.


\section{Observation and data reduction}
The data were collected during four half nights at UT1+FORS2 in High Time Resolution spectroscopy (HIT-S) mode\footnote{http://www.eso.org/instruments/fors1/ for more information.}.
This observing mode allows fast time resolved spectroscopy with little overheads by shifting the
charges on the detector while exposing. Using a mask on which a small slit has been cut, only a small part of the detector get actually exposed to light. The exposed part is then shifted under the masked area while an new spectrum is taken. The sequence continues till the whole CCD has been used-up and it has to be read-out. 
By using a mask with slit 1.0" wide and 10`` long, in 4 nights, we collected 15 frames containing 41 spectra each. 
Each spectrum corresponds to 64~s of integration, hence the total exposure time per frame adds-up to 2624~s ($\sim$44 min).

The adopted instrument setup was grism 600B covering the 3550-6600\AA \ wavelength range. A gap of $\sim$10\AA \ at about 5000\AA \ was present due to the particular orientation of the grism in the HIT mode\footnote{The orientation is 90$^o$ with respect to the standard FORS configuration. Hence, the spectra are no longer parallel to the CCD-mosaic rows but to their columns.}. The resulting spectral resolution was $\sim$5\AA.

  \begin{figure*}
   \centering
  \includegraphics{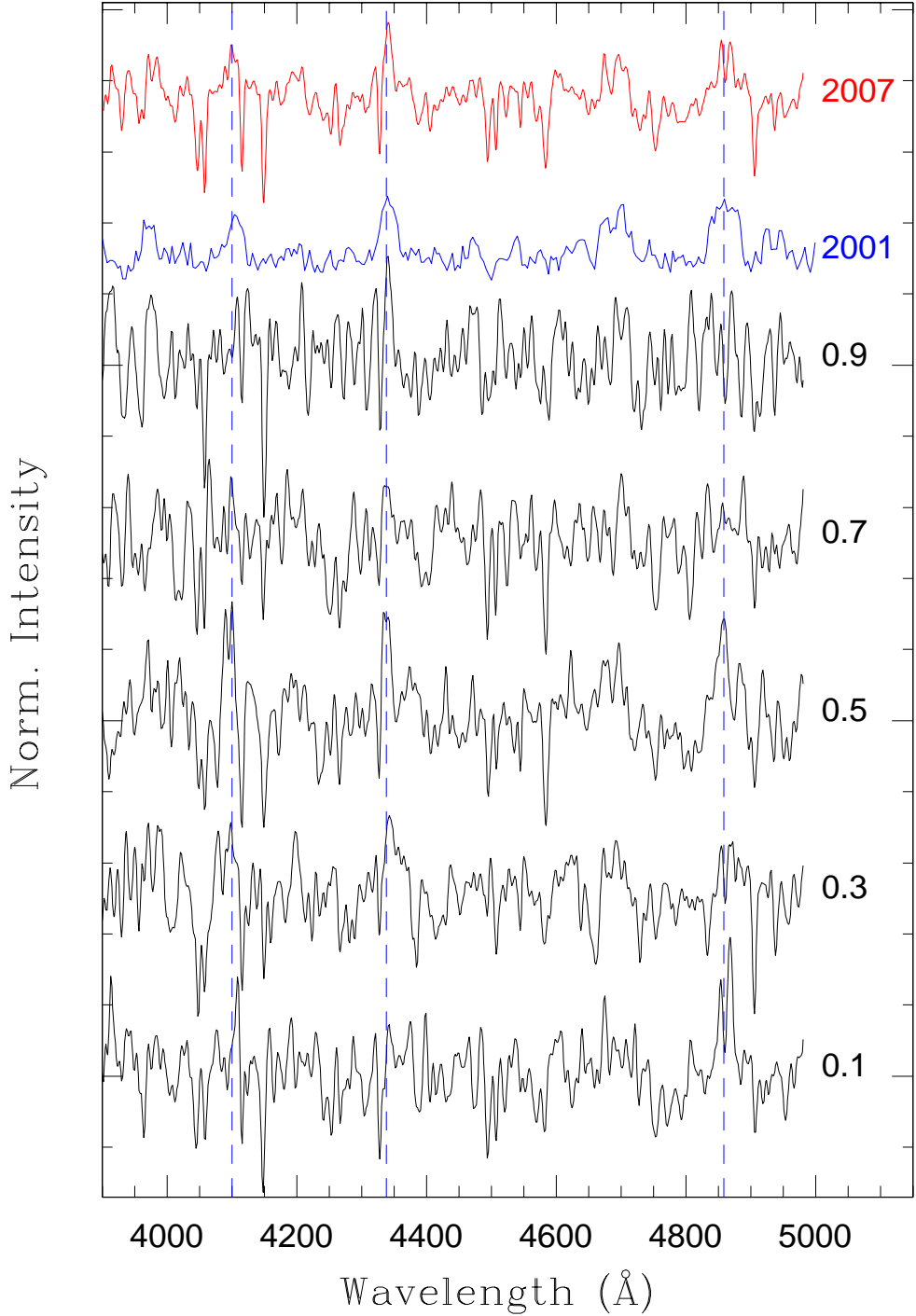}
   \caption{The five 2007 VLT-FORS2 phase-resolved spectra of HM~Cnc are shown plus, at the top, the 2001
FORS1 and 2007 FORS2 phase-averaged spectrum (adapted from I02). All the spectra are normalized to one
and arbitrarily shifted for clarity. The numbers on the left of the spectra mark the center of the 0.2-long phase
interval. Vertical dashed lines mark the HeII 4100.0~\AA, HeII 4338.7~\AA, and HeII 4859.3~\AA. The spectra were re-binned at the spectral resolution imposed by the slit width (4.5~\AA).}
              \label{Fig1}%
    \end{figure*}


The data produced within the FORS HIT-S mode looks similar to multi object spectroscopy data, being they a collection of spectra in a single frame. However, as we experienced during the data reduction process, the handling of HIT-S mode frames requires particular attention.  In particular, we have verified that the standard frame pre-processing (bias subtraction, flat correction etc) is not correct as it introduces noise. 
The HIT-S mode frames present a non uniform  and variable bias level which increases in the direction of the charge shift.  In the majority of our science frames we observed a sloped bias with an increase of $\sim$5-7~ADUs, or 2-4\%, from left to right. In three frames the ''bias slope'' was steeper with the bias level increasing by 14\% or up to 250~ADUs across the detector. 
We discarded those three frames in order to work with an ''homogeneous'' data set and reduced the remaining 'good'  frames subtracting, in place of the bias, a third order polynom fitted along the unexposed rows of the detector. 


We also discarded the flat fields for the following two reasons. First, their showed yet a different and variable ``bias level``. 
Second and most important, the flat field and our science data are characterized by completely different counts level regimes so that the first does not correctly remove the pixel-to-pixel variations of the second.  Due to the relatively high spectral resolution, the faintness of our target and the short exposure time, the signal of the object+sky was only a few ADUs above the bias level. In low count level regimes of 10-50~e$^-$/pix, CCDs suffer from deferred charges or charge traps  (Mc~Lean 1997). Hence, charge traps are most visible in the case of short exposures or spectroscopic observations (for which the dispersed background is below 100e$^-$). Our dispersed signal is of the order of 10e$^-$, only. Delayed release of trapped charges can occur in time ranges from a few second to hours. The net result will be 1-2 pixels wide dark (or bright) short strips in the direction of the charges tansfer and an overall increase of the noise. We note that in our case these features are of the same order of the noise and
therefore are not visible in the single spectra but only in the final median/average spectra for which the noise is significantly reduced. We have verified that this is the case by reducing the same data sets following two different strategies (see below) and comparing the results. We have identified at least 6 CCD rows affected by charge-traps. 
The release of the trapped charges occurs randomly and cannot produce spurious emission features as would not always appear in the same position on the detector. Hence, they are removed when the median of the phase-bins is computed.


We continued the data reduction following two different strategies both as a cross check and to pinpoint the best data reduction procedure. 
In both cases we phased each single spectrum of each frame and defined 5 phase bins centered at phase 0.1, 0.3, 0.5, 0.7 and 0.9 (see below for the absolute time and phase assignment procedure).
In one case we first sliced each frame in 41 subframes. The 2D slices were then stacked and their median computed  in one of the five phase-bins which where eventually extracted. All the spectra within a given slice were shifted to a common value in the
spatial direction, in order to avoid smearing of the signal because of the limited centering accuracy
during the acquisition. The 5 median phase-bin spectra were wavelength calibrated in
2D  before extraction. 

In the second case all the spectra were extracted from the pre-processed 2D frame, after having
build a reference trace function on the median spectrum of the given night. The single extracted spectra were
wavelength calibrated and combined via median in the 5 bins.
In both cases we did not combined all the pre-processed frames but performed a visual inspection
discarding those spectra that had a particularly bad SNR because of the poor sky conditions. As
a result of this selection 75, 79, 75, 73 and 69 spectra were combined in bin 1,2,3,4, and 5, respectively,
out of the original $\sim$123 spectra per phase bin. We verified that this procedure is equivalent
, if not better, to combining all the spectra (good and bad ones) with different weights (e.g. accounting for the slit losses due to variable seeing conditions).

\subsection{Timing and phasing each spectrum}
  \begin{figure*}
   \centering
  \includegraphics[width=14.5cm]{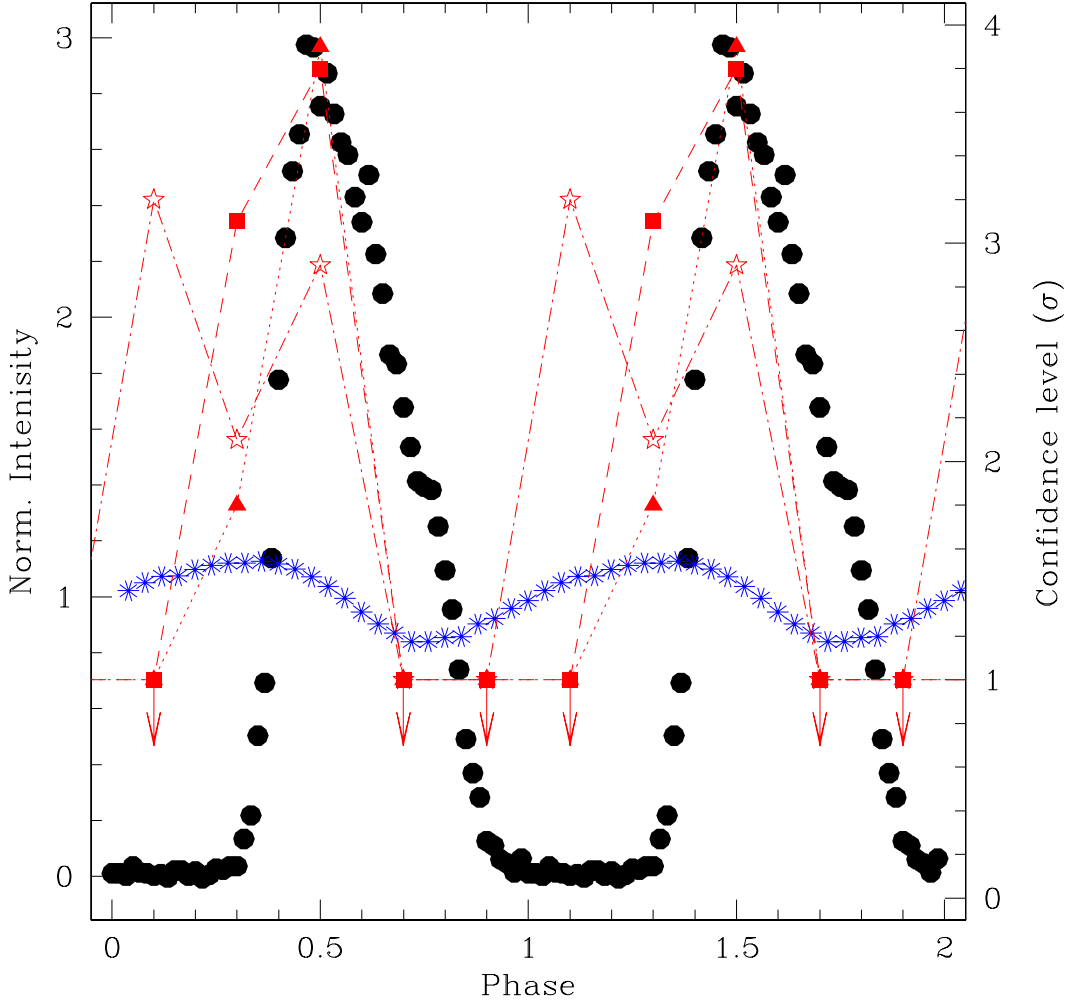}
   \caption{The 1994-2002 phase coherently connected X-ray folded light curves (filled circles; 100\% pulsed
fraction) of HM~Cnc, together with the VLT-TNG 2001-2004 phase connected folded optical light curves
(blue asterisks). For both light curves we have applied the timing solution reported in section 2, after having checked that the arrival times were barycentrized and registered with respect to the same reference system (Terrestrial Time; see also Barros et al. 2007 for a coherent timing solution which includes the covariance terms of the fit ). Two orbital cycles are reported for clarity. The confidence level (in units of $\sigma$) of the three
detected emission lines in the phase resolved VLT-FORS2 spectra are also shown (filled triangles, squares,
and empty stars for 4100.0~\AA, 4338.7~\AA, and 4859.3~\AA \ emission lines, respectively): it is evident the correlation
with the optical pulsed component.}
 
             \label{Fig3}%
    \end{figure*}

Particular attention was payed to the time shift introduced by the (solar system) barycentric correction
which is of the order of few seconds within each night and up to $\sim$30 s ($\sim$0.1 in phase) over
the whole observing run. We computed, a priori, the barycentric correction to be applied to each
frame starting times. This drastically reduced the overlapping phase interval among phase bins (the
overlapping being of order 50\% for the uncorrected frames). The time accuracy of the five phase
bins is of order 20-25\% or $\sim$14 s out of the 64 s exposure time of the spectra. The $\sim$14 s uncertainty
is mainly due to the unfortunate use of 64 s instead of 64.3 s (= 321.53 s/5) as effective exposure
time for spectra.
The ephemeris used for the phase assignment are those of the coherent timing solution by Israel
et al. (2004), and extrapolated to MJD 54144.0 (the first observing night of the run). The period
extrapolated to MJD 54144.0 is P=321.52555(3) s. Note that we did not correct for the P variation
during the observing run baseline, since it accounts for up to a maximum of 2$\times$10$^{-5}$ s within five
nights, well below the 14 s phase accuracy.

\section{Phase-resolved spectroscopy}

Given the relatively poor statistics, we did not attempt to study the continuum of each phase-resolved
spectrum, nor we looked for new and/or unexpected features (whether in emission or in
absorptions). We rather focused on the normalized phase-resolved bin-spectra searching for the strongest emission lines already detected in 2001 (see Figure 1). A visual inspection resulted in the possible
identification of three emission lines at about 4100~\AA, 4340~\AA, and 4860~\AA \ in the spectrum centered
at phase 0.5. 
Therefore we fitted the normalized spectrum with three Gaussians centered at the three wavelengths reported above and measured for each of them the intensity, FWHM, $\lambda_c$, and EW. In the phase 0.5 spectrum the detection of these lines is confirmed at a $\simeq$3$\sigma$ level and the best fit parameter are reported in Table~1. 
By applying the same analysis to the remaining spectra, we obtained only marginal detection ($\sim$2-3$\sigma$ level) of the same lines in the spectrum centered at phase 0.3 and no-detection in the spectra centered at phase 0.7 and 0.9. In the phase 0.1 spectrum we detect only the HeII~$\lambda$4859 at $\sim$3$\sigma$ level (see Table~1). 
At last, we also fitted the same three lines both in the phase-averaged spectum obtained by combining the 5 phase-bin spectra, and  in the VLT-FORS1 2001 spectrum (see I02 for details). While in the 2001 spectrum the detection of the emission lines at 4100, 4338 and 4859~\AA \ is highly significant ($\geq$6$\sigma$), in the case of the average 2007 spectrum the same lines are only marginally detected as expected when combining a small signal with no signal. However, in the average 2007 spectrum we detect at a $\sim$3$\sigma$ confidence level an emission line centered at $\sim$4700~\AA. This was identified with a blend of CIII, NIII and HeII in the 2001 spectrum  (Israel et al. 2002). 


Table~1 reports the best fit parameters of the detected emission lines together with the upper limits on the intensity of the phase-resolved and phase-averaged spectra. 
The upper limits on the intensity of emission lines were obtained in two different ways. We first evaluated the significance of any additional component (each of the three Gaussian) to the fit of the normalized continuum (set to one by definition) applying the F-test. The F-test probabilities are reported in Table~1 in $\sigma$ units. However, we note that Protassov et al. (2002) pointed out that the F-test may be inappropriate in some circumstances and  lead to incorrect estimates of the significance of a feature. In order to further investigate this issue, we ran a Monte Carlo simulation of 100000 spectra with only the continuum model, and the same spectral
coverage and resolution of the spectra we have analyzed. The results of this simulation gave very
similar values to those inferred with the F-test, confirming the confidence levels reported in Table~1.
Note that the above reported confidence levels refers to a search over
the whole spectrum (the number of trial is the number of independent
bins of wavelength); however our emission line search has been
restricted only to the strongest emission lines detected in 2001 with a
number of trial equal to few for each search. Correspondingly, the  the
reported confidence levels are lower limits. Moreover, we can confidentely exclude that the observed line variability is originated by SNR variations (among phase resolved sepctra) due to the intrinsic flux variations (of the order of 10-15\% at maximum) of the source, since the spectrum where we detected most of the lines (see Figure\,1) is at 0.5 in phase, close to the optical minimum (see Figure\,2).    

It is apparent that three emission lines are significantly detected at phase 0.5, marginally detected at phase 0.3 and  undetected in the three remaining phase-resolved spectra (with the exception of the line around HeI $\lambda$4859 at phase 0.1).  The computed upper limits are such that we can confidently conclude that the emission lines in the spectra centered a phase 0.1, 0.7 and 0.9 are either absent or considerably weaker than in the spectra centered at phase 0.3 and 0.5.

\begin{table*}
\caption{Measured parameters of the emission lines detected in the phase resolved VLT-FORS2 spectra of HM~Cnc.}
\label{Tab1}
\centering
\renewcommand{\footnoterule}{}  
\begin{tabular}{lccccccc}
\hline \hline
Line ID & \multicolumn{5}{c}{Phase resolved}    &  \multicolumn{2}{c}{Phase averaged} \\
\AA & 0.0-0.2 & 0.2-0.4 & 0.4-0.6 & 0.6-0.8 & 0.8-1.0 & 2007 & 2001 \\
\hline
HeII & I$<$0.07 & I=0.07$\pm^{0.06}_{0.03}$ & I=0.16$\pm$0.04 & I$<$0.08 & I$<$0.07 & I=0.05$\pm$0.03 & I=0.06$\pm$0.01 \\
4100 &          & FWHM=24$\pm$7 & FWHM=16$\pm$3 &  &  & FWHM=17$\pm^{12}_7$ & FWHM=22$\pm$4 \\
     &  & $\lambda_c$=4087$\pm$6 & $\lambda_c$=4095$\pm$3 &  &  & $\lambda_c$=4097$\pm^5_9$ & $\lambda_c$=4104$\pm$3 \\
     &  & EW=4.2$\pm^{3.8}_{2.1}$ & EW=6.4$\pm$2.0 &  &  & EW=2.1$\pm^{2.0}_{1.6}$ & EW=3.3$\pm$0.8 \\
     &  & [2.8$\sigma$] & [3.9$\sigma$] &  &  & [1.6$\sigma$] & [1.5$\sigma$] \\
\hline
HeII & I$<$0.07 & I=0.11$\pm${0.03} & I=0.18$\pm^{0.05}_{0.04}$& I$<$0.06 & I$<$0.07 & I=0.09$\pm$0.03 & I=0.09$\pm$0.01 \\
4338.7 &          & FWHM=26$\pm$9 & FWHM=13$\pm$3 &  &  & FWHM=29$\pm$8 & FWHM=25$\pm$4 \\
     &  & $\lambda_c$=4348$\pm$4 & $\lambda_c$=4339$\pm$2 &  &  & $\lambda_c$=4336$\pm$3 & $\lambda_c$=4341$\pm$2 \\
     &  & EW=7.2$\pm$3.2 & EW=5.9$\pm^{2.1}_{1.9}$ &  &  & EW=6.5$\pm$2.8 & EW=5.6$\pm$1.1 \\
     &  & [3.1$\sigma$] & [3.8$\sigma$] &  &  & [2.5$\sigma$] & [9.0$\sigma$] \\
\hline
HeII & I$<$0.13$\pm$0.04 & I=0.06$\pm^{0.02}_{0.02}$ & I=0.13$\pm$0.05 & I$<$0.05 & I$<$0.06 & I=0.05$\pm$0.02 & I=0.08$\pm$0.01 \\
4859.3 &   FWHM=26$\pm$9 & FWHM=33$\pm$10 & FWHM=23$\pm^{16}_8$ &  &  & FWHM=32$\pm$12 & FWHM=39$\pm$5 \\
     & $\lambda_c$=4864$\pm$4 & $\lambda_c$=4871$\pm$8 & FWHM=4856$\pm$3 &  &  & $\lambda_c$=4862$\pm$6 & $\lambda_c$=4860$\pm$2 \\
     & EW=8.5$\pm$3.9 & EW=5.0$\pm$2.2 & EW=7.5$\pm^{6.0}_{3.9}$ &  &  & EW=4.0$\pm$2.2 & EW=7.8$\pm$1.4 \\
     & [3.2$\sigma$] & [2.1$\sigma$] & [2.9$\sigma$] &  &  & [2.0$\sigma$] & [8.0$\sigma$] \\
\hline
blend$\sim$4700 & I$<$0.07  & I$<$0.07 & I$<$0.07 & I$<$0.07 & I$<$0.07 & I=0.07$\pm$0.02 & I=0.07$\pm$0.01 \\
 & & & & & & FWHM=15$\pm$4 & FWHM=18.4$\pm$2.5 \\
 & & & & & & $\lambda_c$=4696$\pm$5 & $\lambda_c$=4692.5$\pm$2.5 \\
 & & & & & & EW=2.6$\pm$1.0] & EW=3.2$\pm$0.6 \\
 & & & & & & [3.5$\sigma$] & [7$\sigma$] \\

\hline
\end{tabular}
\\
\scriptsize 
Note -- Emission line intensity (I) are in ADUs; FullWidth Half Maximum (FWHM), central wavelengths ($\lambda_c$),
and equivalent widths (EW) are in units of \AA. Intensity upper limits to emission lines are at 1$\sigma$ confidence
level. The emission line detection level is reported in square bracket in units of $\sigma$ (see text for more details).
All the reported uncertainties are at 1$\sigma$ level.
\end{table*}

\section{Discussion}

The VLT-FORS2 campaign we carried out in February 2007, though affected by poor sky conditions
(seeing in the range 1.2-2'', and thin-thick clouds), allowed us to obtain the first phase-resolved
spectra of HM~Cnc 
and the following two major findings: 
\begin{itemize}
\item[\bf -]{\bf Line variability} The three most intense emission lines identified in the 2001 phase-averaged spectrum
are significantly detected only in spectrum centered at phase 0.5, 
marginally detected in the spectrum at phase 0.3 and absent or considerably weaker in the spectra centered at phase 0.1, 0.7 and 0.9. This implies that the intensity of the emission lines varies across the period.
\item[\bf -]{\bf Line width} We measure narrower FWHMs for the emission lines (particularly in the case of lines HeII lines 4100 and 4338\AA) in the phase resolved spectrum centered at phase 0.5 than in the phase-averaged spectra taken in 2001 and 2007. The most straightforward interpretation is that the emission lines of the phase averaged spectra are Doppler broadened. However, the accuracy of our measurements is not sufficient to be conclusive about this issue. 
\end{itemize}

Additional observational constraints can be derived by comparing the optical and X-ray light curves with our phase resolved spectra. 
In figure~2 we show the X-ray (black filled circles) and the optical (blue crosses) light curves together with the the statistical significance of the detected emission lines (red symbols, different symbols are for different emission lines), as reported in Table 1. 
It is evident that the intensity of the emission lines peaks at the time of the X-ray maximum/pulse. However, the lines are likely present (absent) also at orbital phases when the X-ray pulse is off (on). At the same time, the emission lines region (ELR from now on) and the source of the optical pulse  are both off in the phase interval $\sim$0.6-0.9,  though the ELR stays undetected possibly till phase 1.1.

  \begin{figure}[h]
   \centering
  \includegraphics[width=9cm]{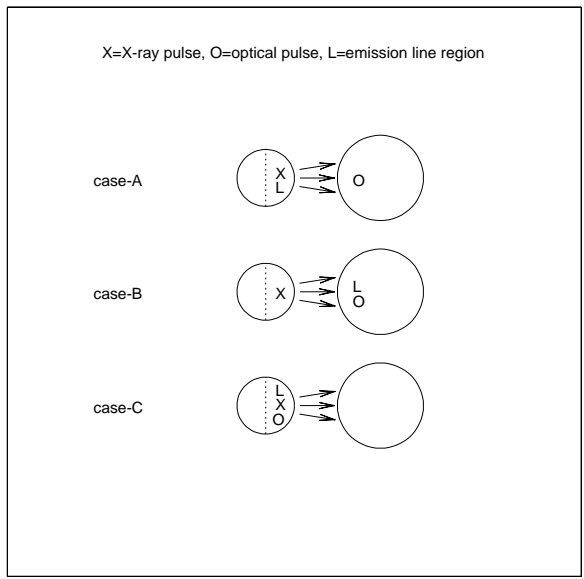}
  \caption{A sketch of the three configurations described and discussed in section 4. The actual position of the labels X,O and L (described in the figure) on the two star components is just indicative. See text for mode details.}
              \label{Fig2}%
    \end{figure}

We can imagine three different scenarios depending on the geometrical location of the X-ray and optical pulse and the ELR (see Fig.\ref{Fig2} for a sketch):\\ 
\noindent {\it \_ case A:} \\
a hot spot on the primary white dwarf is the source of the X-ray pulse and of the optical emission lines, while the secondary star, irradiated by the primary, is the source of the optical pulse (Fig.\ref{Fig2}, top panel). 
This is the most natural configuration if one considers the phase match between the X-ray maximum and the maximum of the emission line intensity as a starting point. The secondary star can then be irradiated either by the X-ray emitting hot-spot or by the hot primary white dwarf. In the former case, the X-ray emitting hot-spot has to be at an angle $\geq$90$^o$ with the line connecting the two stars, in order to be the irradiating source. However, this geometry makes it difficult to account for the phase lag which we observe between the optical and the X-ray pulse (the phase lag should approach 0.5 in phase and not 0.2-0.25, but see also Barros et al. 2007). In the case the irradiating source is the primary WD, on the other hand, there would be no additional constraints on the position of the X-ray hot spot (a part that imposed by the measured phase offset between the X-ray and the optical light curve). Note that, in this latter case, the WD has to be the source of the optical un-pulsed component at temperature is T$\geq$40000~K (Israel et al. 2002), in order to be the irradiating source. 
In this case-A configuration, independently on the irradiating source, the large observed FWHMs of the emission lines can be explained either by Doppler broadening by the rapidly spinning primary, close to the white dwarf break-up velocity (which represents a severe constraint), or by strong turbulence and large thermal velocities of the gas in the hot-spot region. The observed FWHM can be still accounted for by orbital Doppler smearing if the mass ratio is larger than about 1/2. In addition, the early appearance of the ELR with respect to the X-ray pulse poses further constraints without an immediate/obvious explanations. Namely, while temperature gradients in the hot-spot region can be invoked in order to account for this observation, the lower temperature region is located ``down-stream'' or beyond the X-ray emitting (impact) region; contrarily to what typically observed in the case of accretion induced hot-spots (e.g. WZ~Sge in Mason et al 2000). 
However, an impact region with a down stream temperature gradient could be produced by a fast rotating WD as explained by Marsh and Steeghs (2002) in the direct impact model 

\noindent {\it \_ case B:}\\
As in case~A the X-ray hot spot is on the primary star. However, now, both the pulsed optical component and the ELR are on the secondary star (Fig.\ref{Fig2}, middle panel). This configuration is well supported by the measured FWHM of the emission lines, which is consistent with Doppler broadening by the secondary star orbital motion, and by the good agreement between the optical pulse and the visibility interval of the ELR. In this configuration, the phase range 0.6-0.9  match the the time of the secondary inferior conjunction when the irradiated face of the secondary star is not visible to the observer because of the star  orbital motion around the binary system center of mass.
Similarly to case~A, the phase offset between the X-ray and the optical (plus ELR) pulse is better accounted for assuming that irradiation of the secondary star is due to the hot primary and not to the X-ray emitting hot-spot. 
However, within this scenario there is not an obvious explanation for the fact that the ELR and the X-ray pulse reach maximum at the same phase. 

We also note that the un-pulsed optical continuum accounts for up to 85\% of the total observed B band flux, regardless of its origin (primary or secondary star). Therefore, in this cofiguration where the ELR and the optical pulsed component match (they both originate from the irradiated surface of the secondary), the equivalent width of the emission lines, after subtraction of the un-pulsed continuum, would be 10-25~\AA. These values are typical of AM CVn systems and, more in general, cataclysmic variables.

\noindent{\it \_ case~C:}\\
In this latter configuration the X-ray hot-spot, the pulsed optical  component and the ELR are all located on the primary star (Fig.\ref{Fig2}, bottom panel). In order to explain the fact that the optical pulse and the ELR appear before the X-ray pulse we can imagine an impact region with a thermal gradient down stream due to the rapidly rotating underlying white dwarf (as explained by Marsh and Steeghs 2002 in their direct impact accretor model). Alternatively the three components, though all on the primary, might be unrelated and have different origin: e.g. the X-ray pulse could come from the hot-spot, while the optical pulse plus the ELR would come from the primary star side which is irradiated by the hot secondary, in the hypothesis that the latter is the source of the T$\geq$40000~K un-pulsed optical component.  
As in case~A, the large FWHM could be interpreted either with turbulence and high impact or thermal velocities or with a rapidly spinning white dwarf primary. Similarly to case A, the observed FWHM can be still accounted for by orbital Doppler smearing if the mass ratio is larger than about 1/2. Again we consider the rapidly spinning white dwarf a less likely option as the white dwarf would be spinning at a velocity $\sim$1/6 of the break-up velocity.



Unfortunately we are unable to favor one or the other configuration: case~A and C have to be preferred if the phasing of each component is taken as ``driving constraint''; while case~B should be preferred if the measured FWHM ($\sim$1000 km/s) in the phase-resolved spectra is the most stringent requirement. In addition, none of the configuration is fully convincing as case A and C require a peculiar geometry of the hot-spot region, characterized by high thermal velocities and a ``down-stream'' temperature gradient. While case~B  ``fails'' in explaining the phase coincidence of the X-ray and ELR maxima. 
A fourth alternative configuration might locate the ELR and the optical pulse on the curved stream (accretion stream of flux pipe). This  easily accounts for both the early appearance of the optical pulse and ELR with respect to the X-ray emission and for the phase coincidence of the X-ray and ELR maxima. However, none of the current models has yet explored such a possibility, hence the further discussion of the latter configuration is premature.

At last, there are two more observational constraints imposed by the ELR. First the emission lines at $\sim$4700~\AA \ is observed only in the phase-averaged spectra implying that it is very weak but visible across the whole period. Second, the line HeI$\lambda$4859 is visible also at phase 0.1 and at a higher significance than at phase 0.3 and 0.5. 
These two facts are possibly evidence of a complex ELR geometry and/or of  multiple line forming regions.


\section{Summary and conclusions}

The phase resolved spectra of HM~Cnc presented in this paper have shown that its time resolved study is possible and in particular it is feasible with FORS2 HIT-S mode offered at ESO-VLT, though care must be taken during the data reduction and the data analysis phase because of the low count level regime of the spectra and the deferred charges effect.

Our phase resolved sequence has demonstrated the emission lines intensity varies across the period and that the emission lines in the phase-resolved spectra have smaller FWHM than those in the phase-averaged spectra. 
These are two strong observational constraints which however do not allow us -yet- to formulate any definitive conclusion about the location of the line forming region and on the nature of the period because of the low SNR and quality of our data. 
In particular, we cannot establish whether the emission lines arise from the X-ray emitting hot-spot or from the irradiated surface of the secondary star. A higher SNR series of spectra would allow us to 1) establish whether the ELR is visible at any phase or they gets really ``eclipsed''; 2) measure with smaller uncertainties the emission line centroids thus establishing whether they move across the period and the larger FWHM of the phase-averaged spectra is indeed an effect of Doppler broadening. Case~A and C, which locate the ELR on the primary hot-spot should reveal emission lines with no o small Doppler shift. While, on the contrary, our case~B configuration implies significant Doppler shift of the emission lines. 
All the discussed geometries are consistent both with unipolar inductor model and the direct impact accretor as those models do not prescribe any specific location of the ELR in their formulation. The face on-intermediate polar hypothesis, instead is not compatible with the variable emissions line intensity nor with their smaller FWHM in the phase-resolved spectra. Hence it does not fit any of the 3 cases discussed in section~4. In the double degenerate polar model, we expect that the ELR is on the primary and therefore it is consistent only with our configuration A or B. This model should be discarded in case large Doppler shift of the emission lines are measured. 


Finally, our phase-resolved spectral analysis showed that the HeII odd-term series emission lines at 4199.8\AA  \ and 4541.6\AA \  were not detected (at any phase). This is consistent with the EWs inferred from the 2001 phase-averaged spectrum and confirms that the odd-term series lines are fainter than the even-term series ones. The emission line centroids $\lambda_c$ are not yet measured with sufficient accuracy to unambiguously address the issue of the presence of H in the spectrum of HM~Cnc. However, we note that recent theoretical (D'Antona et al. 2006) and observational (Reinsch et al. 2007) studies of HM~Cnc envisage the presence of H in DDB systems.


%

 
\begin{acknowledgements}
      This work was partially supported at OAR through grants from Agenzia Spaziale Italiana (ASI),
Ministero dell'Universita e Ricerca (MUR - COFIN), and Istituto Nazionale di Astrofisica (INAF). EM is grateful to
the ESO Director General for the allocation of the DGDF in support of a science leave in Rome and the INAF-Osservatorio
Astronomico di Roma for their kind hospitality. GLI thanks the ESO/Santiago Visiting Scientist Program for the approval
of a scientific visit in Vitacura during which this paper was completed.
\end{acknowledgements}

\end{document}